\documentclass[sn-mathphys-num]{sn-jnl}

\usepackage{graphicx}%
\usepackage{multirow}%
\usepackage{amsmath,amssymb,amsfonts}%
\usepackage{amsthm}%
\usepackage{mathrsfs}%
\usepackage[title]{appendix}%
\usepackage{xcolor}%
\usepackage{textcomp}%
\usepackage{manyfoot}%
\usepackage{booktabs}%
\usepackage{algorithm}%
\usepackage{algorithmicx}%
\usepackage{algpseudocode}%
\usepackage{listings}%
\usepackage{setspace}
\usepackage{geometry}
\theoremstyle{thmstyleone}%

\theoremstyle{thmstyletwo}%

\theoremstyle{thmstylethree}%

\begin{document}

\title{Mars' Young Sedimentary Rocks: Early thinning, late persistence, diachronous boundaries, and a regional dichotomy}

\author*[1]{\fnm{Madison L.} \sur{Turner}}\email{mturner10@uchicago.edu}

\author[2]{\fnm{Sabrina Y.} \sur{Khan}}\email{syked@jhu.edu}

\author[2]{\fnm{Kevin W.} \sur{Lewis}}\email{klewis@jhu.edu}

\author[3]{\fnm{Axel} \sur{Noblet}}\email{anoblet@uwo.ca}

\author[1]{\fnm{Edwin S.} \sur{Kite}}\email{kite@uchicago.edu}

\affil*[1]{\orgdiv{Department of the Geophysical Sciences}, \orgname{University of Chicago}, \orgaddress{\street{5734 S. Ellis Ave}, \city{Chicago}, \postcode{60637}, \state{Illinois}, \country{USA}}}

\affil[2]{\orgdiv{Department of Earth and Planetary Sciences}, \orgname{The Johns Hopkins University}, \orgaddress{\street{3400 N. Charles St}, \city{Baltimore}, \postcode{21218}, \state{Maryland}, \country{USA}}}

\affil[3]{\orgdiv{Department of Earth Sciences}, \orgname{University of Western Ontario}, \orgaddress{\street{1151 Richmond St}, \city{London}, \postcode{N6A 5B7}, \state{Ontario}, \country{CAN}}}

\singlespacing
\abstract{Mars’ sedimentary rocks record Gyrs of environmental change. New data enable the first global analysis of paleo-environment relevant physical properties of these rocks, including layer thickness and accumulation rate. We find that layer thicknesses of post-3.5~Ga sedimentary rocks across the Martian surface show coherent variations at $\sim$1000~km-scale that are inconsistent with simple volcanic and climatic hypotheses for formation, which are consistent with global compositional homogeneity at orbital scales. These data, in combination with new analyses of outcrop age and total rock volume demonstrate a global decrease in layer thickness that predates the eventual drop off in preserved sedimentary rock volume per Myr. The new constraints confirm a diachronous transition in Mars' global sedimentary rock record while also highlighting a regional dichotomy in young sedimentary rock deposits that has not been quantified before.}


\maketitle

\section*{Introduction}

Layered sedimentary rocks record planetary evolution over billions of years~\cite{peters2022macrostrat}. On Earth,  many attributes of the sedimentary rock record, including layer thickness, depocenter location, and rock volume~(accumulation rates) are sensitive to environmental conditions~\cite{mclennan2019sedimentaryrx}. The same is true of these attributes for Mars' sedimentary rocks~\cite{mclennan2019sedimentaryrx,malin_edgett_mars}. Despite this potential, there have been no studies of the global-scale variations in these parameters in Mars' sedimentary rock record or their variations in time.

 \begin{figure}[H]
 \centering
 \noindent\includegraphics[trim=13cm 3cm 0cm 3cm, clip,width=.7\textwidth]{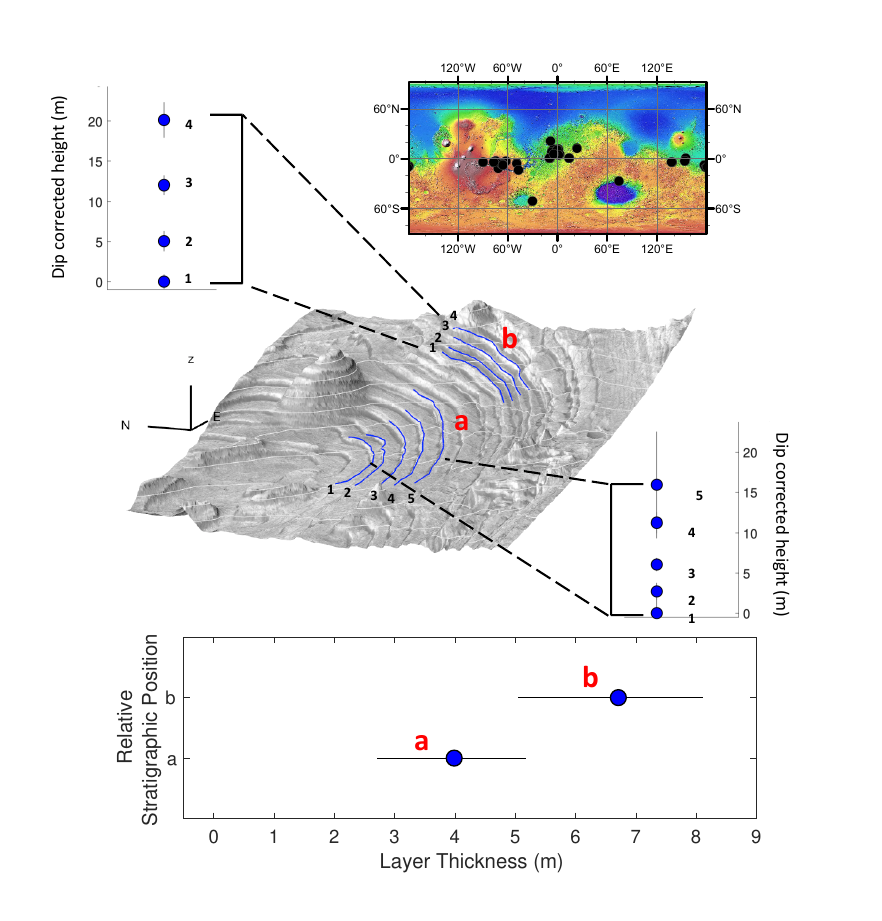}
\caption{Workflow for measuring layer thicknesses.~(Top)~Images and DEMs are obtained from the locations marked in the map~(black markers)~(Fig.~S1).~(Middle)~Sections of continuous layering are identified and traced directly onto the images. Bedding plane orientations are obtained using the layer traces and the corresponding topography. Dip corrected layer thicknesses are obtained for each layer section using the mean layer orientation for the different layer packages~(N:3-11).~(Bottom)~Layer thicknesses are reported as the mean thickness for each package, with the uncertainty quantified using the thinnest and thickest of the individual layer thicknesses. 
}
\label{fig:methods}
\end{figure}

While previous work involved regional studies~\cite{annex2020regional,schmidt2022protracted_arabia_terra,equatorial_layer_deposits_schmidt2021,koeppel2022fragile,dipietro2023groundwater_arabia_terra,annex2023bedding_arabia_terra}, high resolution~(25~cm/pixel) orbital images~\cite{mcewen2007hirise} and digital elevation models~(DEMs)~(1~m/pixel)~\cite{KIRKhirisestereometh} of different deposits allow us to form the most comprehensive and highest resolution dataset to date~\cite{KIRKhirisestereometh,grotz_mill_2012,LewisAharonson2014}, which permits accurate measurements of layer thicknesses at different outcrops within the global sedimentary rock record. This record consists primarily of, in stratigraphic order, the clay, sulfate and rhythmite units, which are distinguished by composition, facies and stratigraphic position~\cite{grotz_mill_2012}. These rock units demonstrate Mars’ environmental transition from a planet with abundant liquid water~(globally distributed clays) to present day Mars, which is cold and dry~(globally distributed anhydrous oxides)~\cite{mclennan2019sedimentaryrx}.

As they comprise our best record of the largest known environmental catastrophe, understanding the formation, growth and evolution of these stratigraphic units is critical~\cite{GROTZmissionobjectives}. Clays on Mars have been hypothesized to record abundant aqueous activity~\cite{malin_edgett_mars,grotz_mill_2012,MILLIKENoggale}, and large deltas and channels visible from orbit formed around the time that the clays formed. 
In contrast, orbital facies of the rhythmites and underlying sulfates consist of laterally extensive, parallel layers of uncertain origin. These units make up the youngest, and most voluminous, portion of the Martian sedimentary rock record. Together, these units apparently accumulated over a time span of $\sim$3.5 billion years~\cite{li2022age_MFF_ages} and account for most of volume of sedimentary rocks on Mars. The change from sulfates to rhythmites has been hypothesized to record global climate change~\cite{bibring2006_geol_time}. Global mechanisms such as ashfall from large caldera-forming eruptions and climate-controlled dust circulation and induration have been proposed for rhythmite accumulation (\cite{kerber2012volcanic,kerber2013volcanic,kite2013snowmeltdust}, but see also ~\cite{mclennan2019sedimentaryrx}). Here we test and reject these hypotheses using new datasets and for the first time synthesize the layer thickness, deposit thickness and volume, and age of Mars' young layered sedimentary rocks. We obtain a surprising new trend where the decrease in sediment rock volume post-dates an initial drop in layer thickness that occurred 3.5 Ga. Based on this result, we suggest a planetary evolution scenario where an initial drop in atmospheric pressure results in thinner layers and a later, gradual decrease in liquid water availability leads to a slow decline in planet-integrated sedimentary rock volume.

\section*{Results and Discussion}
\subsection*{Regional trends in layer thickness, global trends in outcrop age}

We obtained 972 layer measurements from 49 HiRISE Digital Elevation Models~(DEMs) and orthomosaics of sedimentary deposits on Mars~(including new DEMs made for this study) and also analyzed previously collected sedimentary mound volume estimates, along with crater chronology data. The data selection was largely controlled by data availability~(Supplementary Materials), and image pairs were chosen to sample each of Mars' major rhythmite and sulfate deposits~\cite{grotz_mill_2012}. We require accurate layer thickness measurements throughout an outcrop, so mosaics with continuous, laterally consistent and relatively undeformed layering were prioritized.

We obtained data from sedimentary mounds at Valles Marineris, Arabia Terra, Meridiani Planum, the Medussae Fossae Formation~(MFF), in addition to Galle, Terby, and Gale crater~(Fig.~S1).

For the first time, we report large differences in layer thickness between regions: Valles Marineris layers are on average 2$\times$ thinner than the Arabia Terra sites with a maximum layer thickness that is 23m smaller. Furthermore, layer thicknesses are more uniform within Valles Marineris: mean fractional variance of these sites is 0.8$\pm$0.4 versus 2.1$\pm$2.1 for Arabia sites~(Figs.~S13,~S15). Thus, accumulation in these different geographic areas was likely governed by regional depositional processes~(Figs.~4,~S3,~S6,~S11) (although in principle global dust storms could account for this type of variability with a single sediment source~\cite{fenton_duststorms}). Valles Marineris sites show thinning away from the Tharsis volcanoes, as expected for a volcanic origin~\cite{kerber2012volcanic,kerber2013volcanic}, although this trend is slight. Conversely, the Arabia and Medusae Fossae Formation sites lack clear intra-regional trends in latitude or distance from volcanic sources. The differences in trends and in fractional variance between deposits suggests that sediment deposition in these two regions was governed by independent regional controls/independent sediment sources.
Globally we do not observe a trend of thicker layers at lower latitudes~\cite{kite2013snowmeltdust}, nor thinning layers away from the major volcanic centers~\cite{kerber2012volcanic,kerber2013volcanic}. This is inconsistent with previously proposed simple models.

To assess trends with time, crater counting statistics were obtained from the Arabia and Valles Marineris deposits~(Materials and Methods) and used together with existing statistics from Gale crater~\cite{palucis_galecrater,li2022age_MFF_ages}, the MFF~\cite{li2022age_MFF_ages}, Terby crater~\cite{ansan2011stratigraphy_terby}, and Meridiani Planum~\cite{hynek2017geologic_meridiani}. These data place both upper limits on the age of sedimentary rocks~(from the crater retention age of underlying materials) and lower limits on the age of sedimentary rocks (from the crater retention age of the sedimentary rocks themselves), allowing for our data to be ordinated in time. We found that the oldest deposits~(intercrater mounds in Arabia Terra, and Terby crater~(Fig.~4)), have thicker layers than younger regions~(Fig.~2). Moreover, deposits with the same orbital facies have vastly different ages~(Fig.~5), ruling out global climate as the pacemaker of the shift from sulfate orbital facies to rhythmite orbital facies ~\cite{bibring2006_geol_time}.

Using age estimates and mound boundaries~(Materials and Methods), we estimated the total accumulation of sedimentary rock volume as a function of time (i.e., km$^3$/yr)~(Fig.~3). While previous work assumes a gradual drying out of the Martian surface, with a drop-off in sedimentary rock formation with time~\cite{grotz_mill_2012}, figures 2--3 show a disconnect between the drop in layer thickness, which occurs early in Mars’ history, compared to that in sedimentary rock accumulation, which occurs much later~($\sim$0.4~Ga~\cite{li2022age_MFF_ages}). Early on, there is a strong pulse of accumulation and preservation at Meridiani Planum. However, other older deposits~(Gale sulfates, Terby crater, Arabia Terra rhythmites) correspond to a global low in sedimentary rock volume. Sediment volume rebounds with the accumulation of the Valles Marineris mounds, and later, the MFF deposits, before finally dropping off to no sedimentary rock accumulation in the present day. The cause of the apparent secondary rebound in rock accumulation after the low preservation levels immediately following Meridiani Planum is unknown: a similar pattern in terrestrial data corresponds to Earth's Great Unconformity~\cite{peters2022macrostrat}, but other possibilities exist for Mars.

\begin{figure}[H]
\centering\includegraphics[trim=0cm 3cm 0cm 3cm, clip,width=.8\textwidth]{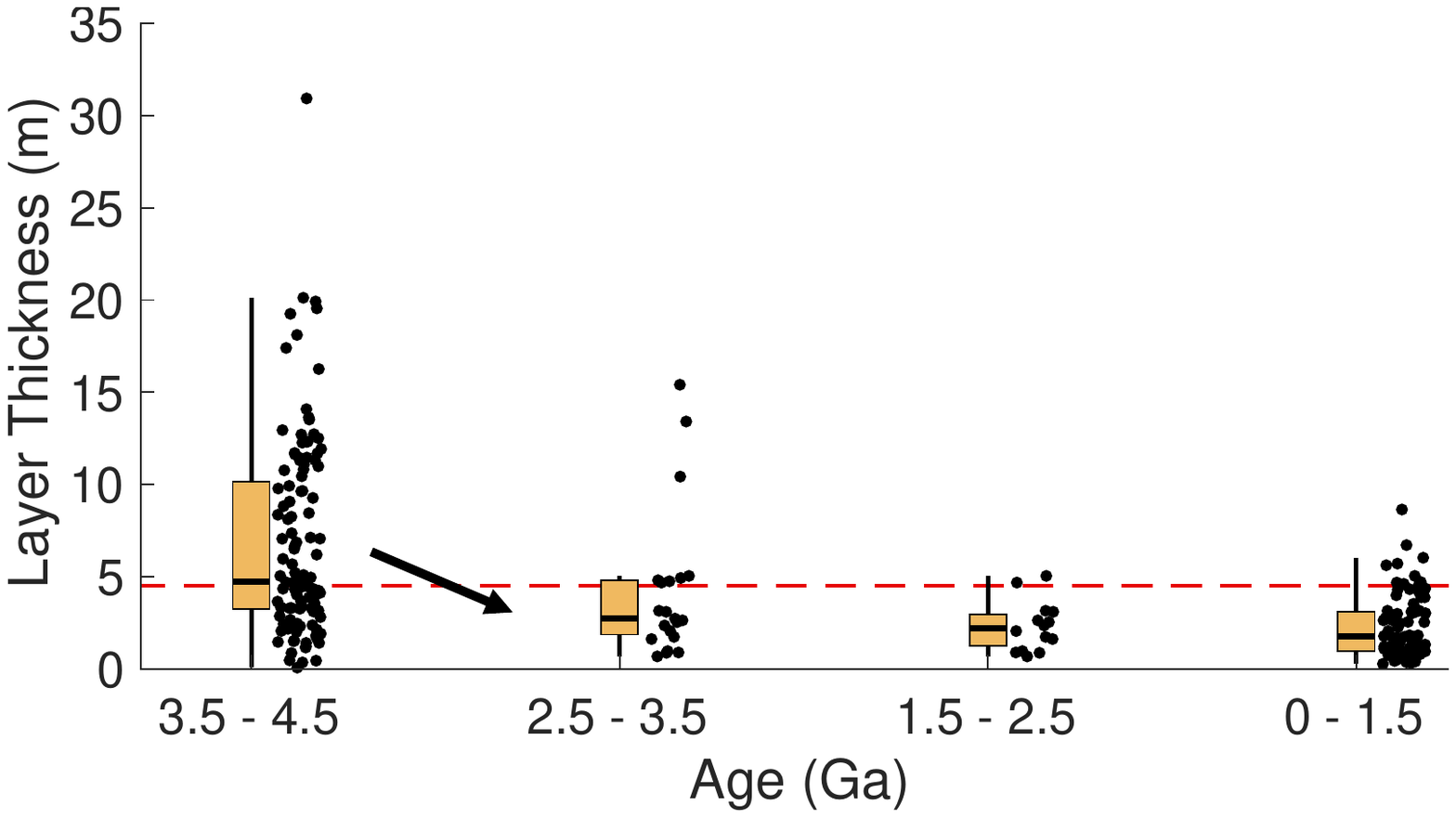}
\caption{Layer thicknesses of measured packages decrease over time. Regions in the 3.5-4.5~Ga bin include sulfates and rhythmites from Meridiani Planum~\cite{hynek2017geologic_meridiani}and sulfates from Gale crater~\cite{palucis_galecrater}. Valles Marineris sulfates and deposits at Terby crater~\cite{ansan2011stratigraphy_terby} are in the 3.5-2.5 Ga bin, and Gale crater rhythmites and the Eastern MFF are in the 2.5-1.5 Ga bin~\cite{li2022age_MFF_ages,zimbelman2012MFF}. The youngest bin contains layer thicknesses from Zephyria Planum and the Valles Marineris rhythmites~\cite{li2022age_MFF_ages}. The width of each box plot is determined by the 25th and 75th percentile value, while the median is plotted with the black line. Whiskers are drawn to show the minimum and maximum values when excluding all outliers. Outliers are defined using the interquartile range (IQR), where a value is defined as an outlier if it is below Q1 - 1.5*IQR or above Q3 + 1.5*IQR. The actual data~(black points) is plotted to the right of each bin. }
\vspace{0cm}
\end{figure}

\subsection*{Rhythmite formation on Mars}

Our new layer thickness data exclude both the simple volcanic~\cite{kerber2013volcanic} and snowmelt-controlled deposition~\cite{kite2013snowmeltdust} models as a global depositional mechanism. The snowmelt model predicts layer thinning at higher latitudes~\cite{kite2013snowmeltdust}, inconsistent with these data~(Fig.~S4). A global volcanic ashfall depositional model is also not consistent with these data because the global minimum in layer thickness in Valles Marineris is closest to Tharsis, Mars' primary magmatic center~(Figs.~S1,~S6). Instead of global control, semivariogram results indicate a regional~(10$^3$~km scale) control of layer thickness~(Fig.~S11).

The spatial variation of layer thickness within the Valles Marineris deposits could be consistent with volcanic ashfall, given the subtle drop-off in layer thickness toward the southeast, away from the Tharsis volcanoes~(Fig.~S9). The layers in these deposits also thicken toward the equator, as predicted by the snowmelt model. However, other outcrops deposited during the same time period lack equatorial thickening. Because of this, we feel that a regional volcanic-source scenario is the more likely of these options. The thicknesses of the Arabia Terra deposits lack intra-regional trends~(Figs.~S4,~S9), meaning that neither of the simple models explain the deposits in this region. This is consistent with prior studies of this region that favor groundwater upwelling as a control on sedimentary rock accumulation ~\cite{schmidt2022protracted_arabia_terra,dipietro2023groundwater_arabia_terra, andrewshanna_lewis_groundwater}, but see also \cite{equatorial_layer_deposits_schmidt2021,koeppel2022fragile}. 

Over time, layers thinned~(Fig.~2) and the amount of sedimentary rock of a given age that has been preserved (km$^3$/yr) decreased over time~(Fig.~3). In principle, these trends could be caused by a decrease over time in sediment availability~(sediment starvation), an increase in erosion, or a decrease in cementation potential~(for conversion of sediment into rock). We now consider each in turn.

The sediment starvation hypothesis predicts that the sediment-rich times correspond to older, thicker layers. However, older rhythmites (yellow wedges at 3.7 Ga in Fig. 3, corresponding to Arabia Terra) formed during a period with a lower volume of rock preservation. Most of Mars' sedimentary rocks are younger~(i.e. Valles Marineris, MFF). The presence of abundant young sedimentary deposits implies that sediment must have been available much later in time (Fig. 3). Alternatively, thinning layers might correspond to an increase in erosion over time. However, this contradicts the expected decrease in erosion potential over time as Mars' atmosphere was lost~\cite{armstrong2005atmosphere_erosion}.  
Therefore, we propose that the decrease in layer thickness and rock volume most likely corresponds to a decrease in cementation potential with time (no sedimentary rock is forming on Mars today). 

 \begin{figure}[t]
 \centering
 \noindent\includegraphics[trim=0cm 3cm 9cm 3cm, clip,width=.8\textwidth]{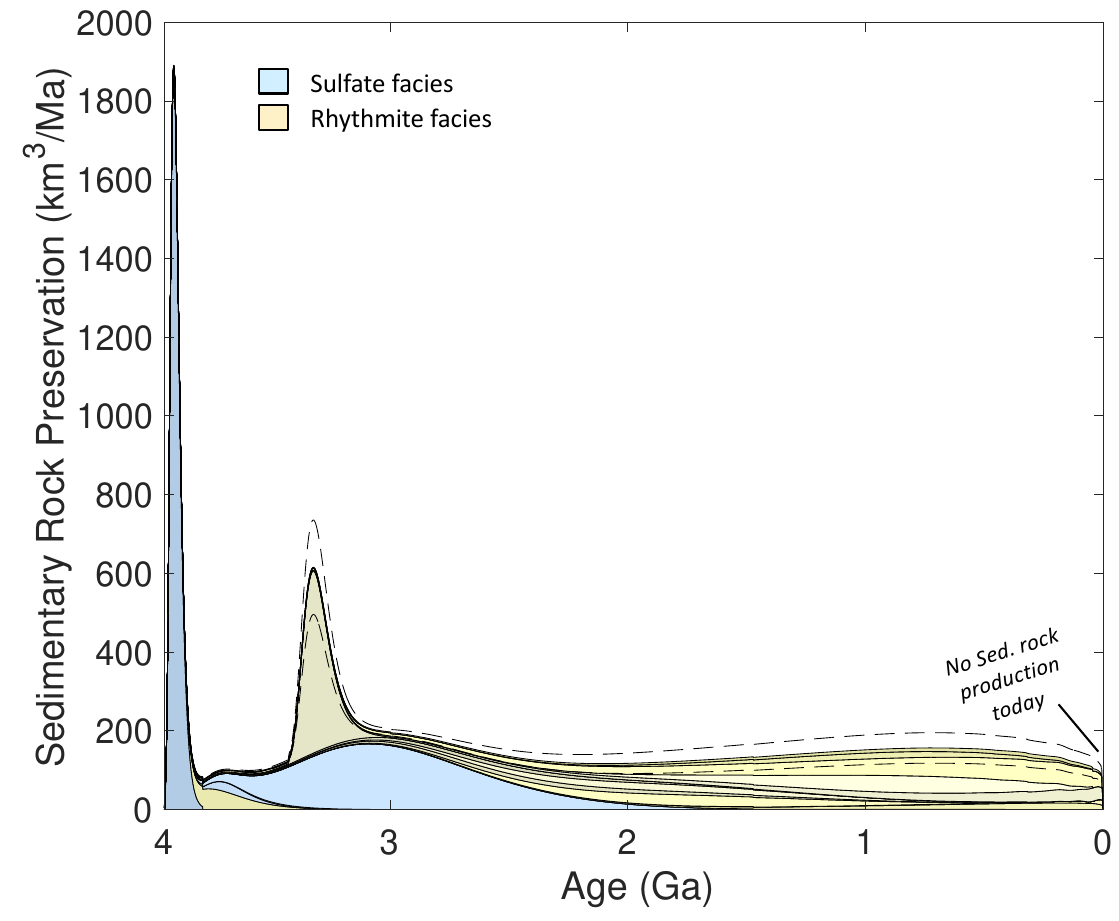}
\caption{Estimate of the minimum rate of sedimentary rock formation through time. Each curve and the underlying shaded area~(blue for sulfates, yellow for rhythmites) marks a different sedimentary deposit mapped on the Martian surface, shown in map view in Figure 4. A detailed breakdown of all mounds represented in this plot is shown in Figure S17, and the uncertainty in MFF mound volumes~(Fig.~S14) is represented by the dashed lines. In most cases, ages are obtained using the crater age of the underlying surface as lower bound and the exhumation age, the age of the current exposure, as an upper bound. All crater retention ages were obtained using a crater counting software, either within this study or previous work~\cite{palucis_galecrater,li2022age_MFF_ages,ansan2011stratigraphy_terby,hynek2017geologic_meridiani,craterstats,robbins_crater_database,michael2013crater_chron}~(Supplementary Information, Table S2). Ages for the Medusae Fossae Formation (MFF) mounds and rhythmite deposits in Valles Marineris are formation age estimates corrected for erosion using the methods of ref. ~\cite{li2022age_MFF_ages}. 
}
\label{fig:age_vs_vol}
\end{figure}

Though a decrease in cementation potential can explain decrease in rock formation rate and layer thickness over time, it does not explain the decoupling in the timing of these declines: layer thickness by a factor of $\sim$2 $\sim$3~Gyr~(Figs.~3-4) before sedimentary rock formation rate declines. 
We propose that this can be explained by an initial drop in atmospheric pressure, which would inhibit erosion~\cite{armstrong2005atmosphere_erosion}, prior to the start of decreasing cementation potential caused by loss of liquid surface water~(Fig.~5). The oldest thickest layers~(Arabia Terra rhythmites, Gale and Terby sulfates) would be deposited before the drop in atmospheric pressure, leading to a low preservation potential due to erosion by a thicker atmosphere
The persistence of sedimentary rock volume late into the Amazonian signals a drop in atmospheric pressure allowing for sedimentary rock to be more easily preserved.
The combined presence of a lower erosional potential with a longer-lived, but decreasing, liquid surface water reservoir would then explain the large volume of young sedimentary rocks. The large preserved rock volume of Meridiani Planum sulfates~(Fig. 3) is an outlier relative to the framework described above. We suggest this is explained by greater erosional resistence, implying that these deposits are more indurated. Meridiani Planum sulfates have a large density of craters~\cite{hynek2017geologic_meridiani}, which indicates greater erosional resistance because erosion effaces craters.
\subsection*{Spatial trends in sedimentary rock deposition through time}
Surprisingly, given the complete lack of evidence for continental drift on Mars, we find that late stage sedimentation was unsteady and regularly shifted locations~(Figs.~4-5). Sedimentation stops early near 0$^\circ$ longitude while deposition of rhythmites near 60$^\circ$ W and 150$^\circ$ E persists to $\sim$~0.4~Ga~\cite{li2022age_MFF_ages}. This order-of-magnitude age difference corresponds to the same orbital facies: these deposits all display the characteristic, light-toned, regular layering of the rhythmite facies and always superpose sulfate facies when both units co-occur~\cite{grotz_mill_2012,LewisAharonson2014,hynek2017geologic_meridiani}.
 \begin{figure}[ht]
 \noindent\includegraphics[trim=2cm 3cm 2.5cm 3cm, clip,width=\textwidth]{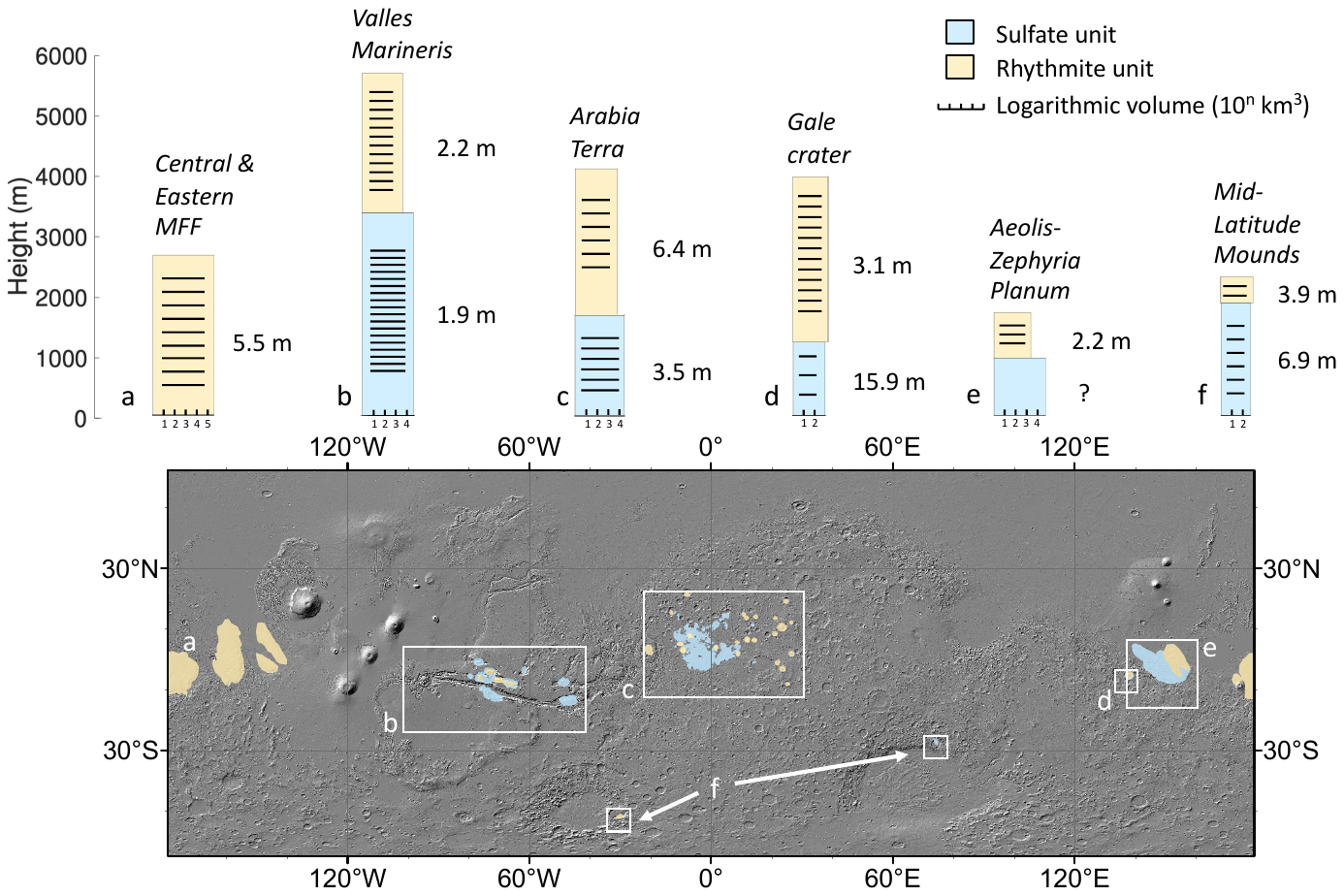}
\caption{Global distribution of the sedimentary rock units analyzed in this study. The representative columns from each region show the relative regional volume~(column width), maximum outcrop thickness~(column height), and layer thickness~(hatching: quantified to the right of each column) of the sulfate and rhythmite units. Mound volumes range over three orders of magnitude. Tick marks at the base of each column correspond to log$_{10}$ units.} 
\label{fig:figure4}
\end{figure}
Thus, orbital facies are not a reliable time-marker in the Mars geologic record (contrary to ref. \cite{bibring2006_geol_time}, but consistent with ref. \cite{mclennan2019sedimentaryrx}). Why did these shifts occur? Possible controls include depositional basin size and age, groundwater table fluctuations or a shift from groundwater to surface water as the water source for aqueous cementation~\cite{andrewshanna_lewis_groundwater,kite2013snowmeltdust}, true polar wander and corresponding shifts in topography~\cite{true_polar_wander_bouley2016,perron2007_truepolarwander_topography}, or the timing of volcanic eruptions~\cite{kerber2013volcanic}. We now address each in turn.

Most mounds analyzed in this study sit within  natural geologic containers. Examples include crater basins and canyon interiors, for which the craters' rims and canyons' walls limit the extent of the interior sediment. The inter-crater deposits in Arabia Terra (which have thicker layers) have much smaller geologic containers than the mounds within the canyons of Valles Marineris (which have thinner layers). This is consistent with a simple model that predicts that the deposits in smaller containers have thicker layers~(Fig.~S12).

When sedimentation started, some containers did not yet exist. Estimates of the age of the Valles Marineris rift system converge around the Noachian-Hesperian boundary~($\sim$3.7~Ga)~\cite{andrewshanna_valles_1}, so lack of basins could have limited early deposition in Valles Marineris. This alone does not explain why sedimentation stopped in Arabia Terra, and does not account for the Medusae Fossae Formation, which has no container.

Groundwater hydrology models predict upwelling in most of the locations where sedimentary mounds occur~\cite{andrewshanna_lewis_groundwater}. If groundwater were the primary source of aqueous ions for minerals in these sedimentary rocks, the spatial shifts could be explained by fluctuations in the groundwater table through time. One model does predict that groundwater flow at Arabia Terra would dry up first~(Fig.~S6)~\cite{andrewshanna_lewis_groundwater}, consistent with our data. 
Another potential mechanism could be a shift in the source of aqueous fluid needed for cementing minerals~(cementation source) from bottom-up~(groundwater controlled), to top-down, snowmelt-supplied~\cite{kite2013snowmeltdust,niles_michalski2012atmospheric}. For this scenario, snowmelt would periodically permit the cementation of wind-blown deposits, forming most of the sedimentary mounds we observe today.

Another way to explain the spatial shifts in depocenters with time is true polar wander. As lava piles up at Tharsis, the planet's axial tilt changes, moving the whole mass equatorward~\cite{true_polar_wander_bouley2016}. This shift would also result in changing topography relative to latitude~(and wind patterns)~\cite{perron2007_truepolarwander_topography}, which affects sedimentation. However, published models predicting different depocenters in these scenarios do not predict surface water accumulation in Arabia Terra, in tension with the data~\cite{true_polar_wander_bouley2016}. Furthermore, because Arabia Terra lies $\sim$90$^\circ$ longitudinally from the Tharsis load, the region's latitude would not change much from true polar wander.

The spatial shifts in depocenter could (in principle) be explained by shifting eruptive centers. Predicted ashfall from calderas overlaps with mapped sedimentary deposits~\cite{kerber2012volcanic,kerber2013volcanic,michalski2013arabiasupervolcanoes}. However, if deposits correspond to direct deposition of ash, then their distribution should match the timing of volcanic eruptions. This does not appear to be the case for deposits in Arabia Terra, which are far from large volcanoes and receive relatively little ash based on atmospheric models~\cite{kerber2012volcanic,kerber2013volcanic,michalski2013arabiasupervolcanoes}.
Age dating of the largest caldera upwind of Arabia Terra~(Nili Patera) shows that eruptions continued long after sedimentation in Arabia Terra stopped~\cite{robbins2011volcanicages}. Because of this mismatch in timing, shifting eruptive centers cannot straightforwardly explain these data.

In addition to variable spatial trends within the same unit, we find that the spatial variability of the rhythmite unit is completely independent of trends in the sulfate unit. Wherever they co-occur, rhythmite outcrop overlies the sulfates. However, despite this global consistency, there are many differences between regions~(Fig.~4). 
The regions with the thickest rhythmites are not the same as the regions with the thickest sulfates. This applies both to absolute thickness of layers, and relative layer thickness compared with other deposits of the same facies. For example, the thickest rhythmite layers in this dataset~(Arabia Terra) are 8,000~km away from the thickest sulfate layers~(Gale crater). Furthermore, previous studies record even thicker sulfate layers in Juventae Chasma within Valles Marineris~\cite{LewisAharonson2014}. This decoupling of spatial patterns between the two facies suggests that the rhythmites and sulfates had independent sources and depositional settings~\cite{fenton_duststorms}. 

 \begin{figure}[t]
 \noindent\includegraphics[trim=0cm 3cm 1cm 3.5cm, clip,width=\textwidth]{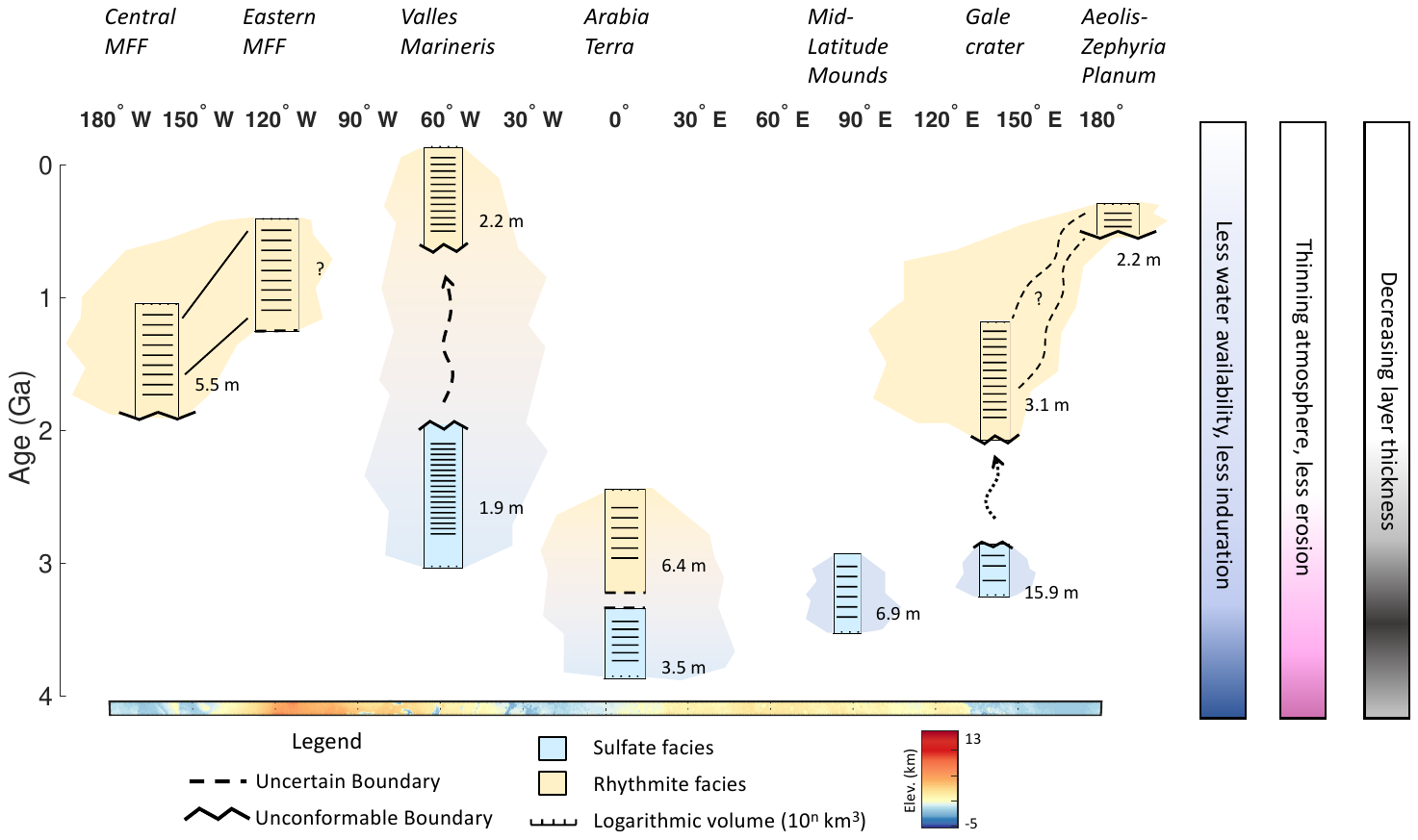}
\caption{Macrostratigraphy of Mars 3.5~Ga to present. Deposit columns and interpreted stratigraphic correlations organized by relative age and longitude. Stratigraphic correlations are marked by arrows and lines and by the colored masks. We find that Valles Marineris and Arabia Terra sedimentary rocks evolved asynchronously despite the similarities in stratigraphy. In contrast, deposits of the Eastern and Central Medusae Fossae Formation, Aeolis-Zephyria Planum, and Gale crater are hypothesized to be correlative~\cite{zimbelman2012MFF}, which would span roughly 120$^\circ$ in longitude around the planet. Alternatively, Gale crater could have evolved asynchronously from the rest of the Medusae Fossae Formation, perhaps via atmospheric teleconnections to Arabia Terra or Valles Marineris. The bars on the right side of the figure display the interpreted changing controls~(cementation potential and atmosphere) and the resulting observable characteristic~(layer thickness). Layer thickness peaks during the deposition of rhythmites in Arabia Terra, and sulfates at the Mid-Latitudes and Gale- in our interpretation, this is after the atmosphere has thinned and before significant loss of liquid water has occurred.
}
\label{fig:figure5}
\end{figure}

\subsection*{An irregular record of layer thickness and sedimentation for Mars}\label{sec13}

These results show that models of Mars’ sedimentary rock formation need to be updated, and are also a benchmark for tests of current and future hypotheses. Our global survey results are the largest, most detailed, and uniform dataset of layer thickness measurements on Mars. We find several new constraints for Mars’ sedimentary rock record: 1) Layer thickness results support provincial, rather than global, mechanisms of transport, accumulation and  preservation~(most notable in Valles Marineris and Arabia Terra)~(Figs. 4, 5); 2) we confirm a diachronous boundary between sulfates and rhythmites on Mars despite consistent facies, composition and stratigraphic ordering~(Figs. 3, 5); and 3) these constraints persist all while layering notably thins post-3.5 Ga~(Fig. 2).

Although layers thin over Gyr timescales, we do not observe any systematic thinning within the stratigraphy of a single DEM~(~10s of Myrs~\cite{lewis_milank_mars_2008})~(Fig. S7). 

The spatial variations in depocenter and layer thickness are robust, and reject simple global models that were built upon observations of global consistency in morphology and composition from orbit \cite{bibring2006_geol_time}. These trends, taken together with diachroneity~(Fig. 5), imply a paleoenvironment that facilitated a dynamic and migratory sedimentary rock cycle despite the absence of plate tectonics.

\section*{Methods}\label{sec11}
Though HiRISE stereopair coverage is sparse, it includes many layered sedimentary mounds that are observed girdling the equator of Mars. We only study mounds previously identified: sedimentary mounds within the Valles Marineris canyon system~\cite{roach2010diagenetic_valles_marineris,wendt2011sulfates_ophir_chasma,okubo2014westcandor}, the layered inter-crater mounds at Arabia Terra~\cite{annex2020regional,annex2023bedding_arabia_terra,dipietro2023groundwater_arabia_terra,equatorial_layer_deposits_schmidt2021,schmidt2022protracted_arabia_terra} and Meridiani Planum~\cite{hynek2017geologic_meridiani}, Mount Sharp within Gale crater~\cite{grotz_mill_2012,LEDEITdipsatgale,thomson2011gale}, the Medusae Fossae Formation~\cite{grotz_mill_2012}, the terraced outcrop in Terby crater~\cite{ansan2011stratigraphy_terby}, and the intercrater mound in Galle~\cite{LewisAharonson2014}. We obtained layer thicknesses with known stratigraphic ordering in at least one HiRISE mosaic and DEM from each of these locations. Each mosaic was assigned to either the rhythmite or sulfate orbital facies to obtain the averages shown in Figure 4. 

We used 49 HiRISE mosaics and DEMs from these outcrops. We only use HiRISE mosaics due to their high spatial resolution (up to 25~cm/pixel imagery and 1~m/pixel DEM). This resolution is necessary given the layer thickness for these deposits; which ranges from sub-1 meter to greater than 10 meters. Because this study uses layering presumably accumulated over orbital timescales~\cite{lewis_milank_mars_2008,LewisAharonson2014}, as a proxy for each orbitally defined sedimentation event, we measure the thinnest, laterally continuous layers that  can be resolved at each location. Data from Curiosity suggests that layering observable from orbit is frequently composed of finer scale, sub-meter, and millimeter-scale laminae. We cannot resolve these smaller features, and purposefully do not measure finer scale sub-layering within orbital outcrops that is not continuously visible. The selection of this subset of DEMs is controlled by data availability and layer exposure. Our dataset includes previously published and publicly available DEMs from existing stereopairs and 7 new DEMs produced from existing stereopairs using the Integrated Software for Imagers and Spectrometers and Ames Stereo Pipeline (ASP) software~\cite{ASP_beyer}. The vertical resolution of the resulting DEMs depends on the horizontal resolution and the convergence angle of the two images in the stereopair~\cite{KIRKhirisestereometh}, and is typically 10 to 30 centimeters. This vertical precision is therefore well below the minimal thickness of the individual layers and layer packages that we measured, as demonstrated in Figure 1. New DEMs were selected based on layer exposure and perceived data gaps, assessed in part based on previous literature~\cite{annex2020regional,annex2023bedding_arabia_terra,dipietro2023groundwater_arabia_terra,equatorial_layer_deposits_schmidt2021,schmidt2022protracted_arabia_terra,LewisAharonson2014,stack_orbital_bed_thicknesses}. Layer exposure also influenced our choice of DEMs. For this work we need accurate, dip corrected thicknesses at different intervals within an orbitally-resolved outcrop. This requires that multiple exposures of laterally continuous and coherent packages of layering be visible. We prioritize layer exposures that lack intense faulting. We also sought to have at least two mosaics for each sedimentary mound to test the null hypothesis that layer thicknesses are spatially uncorrelated. 

Outcrops were assigned to the rhythmite or sulfate facies as follows. If outcrops were mentioned in pre-existing, geologic mapping-based literature as being part of one of the facies, we adopted that assignment. This applied to much of the Valles Marineris canyon system~\cite{wendt2011sulfates_ophir_chasma,okubo2014westcandor}, Meridiani Planum~\cite{hynek2017geologic_meridiani}, the Arabia Terra mounds~\cite{annex2020regional,annex2023bedding_arabia_terra,dipietro2023groundwater_arabia_terra,equatorial_layer_deposits_schmidt2021,schmidt2022protracted_arabia_terra}, Gale crater~\cite{LEDEITdipsatgale,thomson2011gale}, and Terby crater~\cite{ansan2011stratigraphy_terby}. We made new assignments for the sedimentary mounds in East Candor, which has not been extensively mapped since the Viking era, and the MFF, which is mostly dust covered. In these cases, we classified the different units using the orbital facies visible from HiRISE and CTX, underlying and overlying contacts, and spectroscopic indices from the Compact Reconnaissance Imaging Spectrometer for Mars~(CRISM)~\cite{murchie2007CRISM}, using criteria outlined in ref. \cite{grotz_mill_2012}. 

Boundaries between the rhythmite and sulfate units were mapped in Ophir Chasma and Candor Chasma within Valles Marineris~(Fig.~S10). These boundaries were determined based on previous work mapping sulfates~\cite{roach2010diagenetic_valles_marineris}, along with detailed geologic maps of individual mounds that isolate different facies consistent with the rhythmite unit~\cite{wendt2011sulfates_ophir_chasma,lucchitta2015westcandor}. In West Candor and Ophir Chasma, the rhythmite occurs above a large angular unconformity~\cite{wendt2011sulfates_ophir_chasma,lucchitta2015westcandor}. For East Candor we mapped the boundaries based on compositional signatures from CRISM and the overlap with the characteristic orbital facies in HiRISE~\cite{grotz_mill_2012}. The sulfate unit orbital facies consists of laterally consistent~(10s--100s~kms) bedding of variable thicknesses, ranging from $\sim$1 to $\sim$100 meters, with sulfate signatures identifiable from orbit. By contrast, the rhythmite orbital facies consists of light-toned, highly regular and rhythmically layered, fine grained strata with no spectroscopic evidence for aqueous minerals~\cite{grotz_mill_2012}.

Layer thicknesses at each site are determined by first measuring a series of layers throughout each DEM. Within the images, continuous packages of at least 3 layers are identified. The relative stratigraphic ordering of packages is determined using a combination of layer dips, topography, and existing geologic maps~\cite{okubo2014westcandor}. Within each package, the layers are traced manually on the HiRISE image, and the underlying topography is extracted in order to calculate layer orientation. A plane is fit to each three-dimensional topographic trace using Ordinary Least Squares~(OLS) Regression, assuming that all of the uncertainty is confined to the z-direction~\cite{LewisAharonson2014}. The dip and azimuth is calculated from each planar fit, and the associated error is defined as the angular error on the normal to each fitted plane. Planes with an error greater than 2$^\circ$ are considered to not be good fits, and these layers are not considered in subsequent steps when calculating orientations for each package. This 2$^\circ$ is an empirical threshold used in many previous studies measuring orbital layering from HiRISE~\cite{lewis_milank_mars_2008,LewisAharonson2014,annex2020regional,annex2023bedding_arabia_terra}. For this study, 96\% of all measurements satisfy this error threshold. 

Layer fits with large error can be caused by a lack of natural outcrop curvature, manifesting on the surface as a linear exposure without a large enough profile in the xy-plane, or by layer curvature, which can be caused by deformation, block displacement, erosion and slumping. For each continuous layer package, fits with better than 2$^\circ$ error were used to calculate a mean dip and azimuth for the entire package of layering. If the layers measured within the exposure had more than one layer with an error exceeding this threshold, the package was not used, and an alternate location was found instead.

Layer thicknesses are calculated by transforming each layer into the new coordinate space defined by the plane of the mean dip, where the z’ direction, corresponding to height within stratigraphy, is parallel to the normal of mean layer fit for each package. Within this rotated coordinate space, the stratigraphic height of each layer is determined by the mean height of the entire rotated trace, with the uncertainty of each fit shown as the amplitude of the residuals on each layer fit, as shown in Figure 1. Uncertainty in the stratigraphic elevation of the fit has similar causes to the error on the fit itself. We use the mean layer thickness of each package as opposed to individual layer thicknesses for our analyses. This minimizes the error associated with measuring a single layer, because the residuals of a single fit are unlikely to extend the span of an entire layer package given the error constraints on the planar fit for the layers. In addition to reporting the mean, we also show the thinnest and thickest individual layer thicknesses~(Figs.~1,~S13).
 
Between 3 and 11 separate packages of layers are measured within each mosaic. This depends on the availability of measurable, continuous exposed layers. For analyses of spatial trends in layer thickness~(Figs.~4,~S6), the mean layer thickness of all package averages is used as the representative value for a DEM. For analyses of layer thickness with age~(Fig.~2), individual package averages are used.

We also obtain estimates of age and volume for the different sedimentary mounds in this study~(Fig.~S2,~Table~S2). For age estimates, we combined existing age estimates from the literature and our own crater counts. Crater counting and statistics were performed on the mounds in Valles Marineris, the mounds in Arabia Terra~(excluding the Meridiani Planum plateau), and Galle crater. In each of these regions, craters with diameters larger than 2~km, 5~km, and 16~km were identified using the Robbins crater database~\cite{robbins_crater_database}. The diameter thresholds were chosen for consistency with existing crater counts in Meridiani Planum performed by ref. ~\cite{hynek2017geologic_meridiani}. Age estimates were created from the mapped craters using CraterStats~\cite{craterstats}. We used previous age estimates for Gale crater~\cite{li2022age_MFF_ages,palucis_galecrater}, the Medusae Fossae Formation~\cite{li2022age_MFF_ages}, Meridiani Planum~\cite{hynek2017geologic_meridiani}, East Candor~\cite{li2022age_MFF_ages}, and Terby crater~\cite{ansan2011stratigraphy_terby}.

For most mounds, the age range was determined by the age of the underlying surface (i.e. crater floor, canyon rifting), and the exhumation age of the mound. The likelihood of sedimentation was assigned a uniform probability of occurring between the age limits defined by these two bounding ages, and the error on each end was taken directly from the associated age distributions. In some cases, this method assumes that the underlying basin surface has been exposed for the majority of its geologic history, implying that the overlying deposits were not part of larger continuous deposits that have been eroded back. At Arabia Terra, we use the same Noachian aged basement as the underlying strata for the sulfate and rhythmites in Arabia Terra, even though it is speculated that the rhythmites may lie unconformably above the sulfate units of Meridiani Planum~\cite{hynek2017geologic_meridiani}. For Gale crater, we use the exhumation age of the entire mound as the top-age estimate, and the age of Gale itself, believed to be 3.6 Gyrs old~\cite{palucis_galecrater}, for the underlying bounding age. Ages obtained from ref. \cite{li2022age_MFF_ages} are estimates of the entire mound accumulation as opposed to exhumation and basin age. For these locations, we implemented the the age and error distribution directly from the model used for the source study.

Volume estimates were calculated in ArcGIS by interpolating basal surfaces from the mapped unit boundaries and largely follow ref. \cite{tutolo_abstract}. For each boundary, the MOLA topography~(MEGDR 128~pixels per degree)~\cite{smith2001MOLA} was extracted and used to interpolate a basal surface for that unit using Inverse Distance Weighted~(IDW) interpolation. The volume, and maximum height of the units was then determined by subtracting the modern MOLA topography from the interpolated basal surface. For locations with rhythmite and sulfate unit exposures, such as Gale, the rhythmite portion was determined using the method above, and the sulfate portion was determined by subtracting the rhythmite volume from the volume of the entire mound. 

Volume estimates are uncertain, especially for the Medusae Fossae Formation. This uncertainty is rooted in differences in mapping approaches between investigators, small variations in mapped boundaries, and, because this method requires interpolation of a basal surface, the interpolation method used. These uncertainties are biggest for the MFF for several reasons, including dust cover which hinders mapping by obscuring spectral signatures and primary orbital morphology. This uncertainty is demonstrated in Figure S14. Because there are few ground-truth-based constraints on rock volume in the MFF, due to this sensitivity, we estimate the volumes for the MFF mounds have an uncertainty of roughly 50\%, which is represented in Figure 3 by the dashed lines. \\

\noindent \textbf{Data Availability.} Data including images, DEMs, spreadsheets of layer locations and orientations, and shapefiles of different mounds is located in Zenodo (DOI:10.5281/zenodo.13384762).

\bmhead{Acknowledgements}

We acknowledge funding from the NASA Mars Data Analysis Program~(MDAP)~(80NSSC22K1084). We acknowledge David P. Mayer, and Jonathan Sneed for contributing to the calculation of mound volumes, and Dr. Andrew A. Annex for partial creation of HiRISE DEMs used in this study.

\bmhead{Author contributions}
E.S.K. and K.W.L. designed the research; A.N. and S.Y.K. contributed new data; M.L.T. carried out research with assistance from E.S.K.; M.L.T. wrote the paper with assistance from E.S.K. All authors discussed the results and reviewed the manuscript.

\bmhead{Competing interests} The authors declare no competing interests.

\bibliography{main_arXiv_submission}

\end{document}